\documentclass[aps,twocolumn,pra]{revtex4-1}%
\usepackage{amsfonts}
\usepackage{amsmath}
\usepackage{amssymb}
\usepackage{graphicx}
\usepackage{times}
\usepackage{dsfont}
\usepackage[colorlinks=true,citecolor=blue,urlcolor=blue]{hyperref}%
\providecommand{\U}[1]{\protect\rule{.1in}{.1in}}
\begin{document}
\title{ Moiré Superradiance in Cavity Quantum Electrodynamics with Quantum
Atom Gas }
\author{Lu Zhou$^{1,5}$}
\email{lzhou@phy.ecnu.edu.cn}
\author{Zheng-Chun Li$^{1,2}$}
\author{Keye Zhang$^{1}$}
\author{Zhihao Lan$^{4}$}
\author{Alessio Celi$^{5}$}
\email{alessio.celi@uab.cat}
\author{Weiping Zhang$^{2,3,6,7}$}
\email{wpz@sjtu.edu.cn}
\affiliation{$^{1}${State Key Laboratory of Precision Spectroscopy, Department of Physics,
School of Physics and Electronic Science, East China Normal University,
Shanghai 200241, China}}
\affiliation{$^{2}${School of Physics and Astronomy, Shanghai Jiao Tong University,
Shanghai 200240, China}}
\affiliation{$^{3}${Tsung-Dao Lee institute, Shanghai Jiao Tong University, Shanghai
200240, China}}
\affiliation{$^{4}${Department of Electronic and Electrical Engineering, University College
London, Torrington Place, London WC1E 7JE, United Kingdom}}
\affiliation{$^{5}${Departament de Física, Universitat Aut\`onoma de Barcelona,
E-08193 Bellaterra, Spain}}
\affiliation{$^{6}${Shanghai Branch, Hefei National Laboratory, Shanghai 201315, China}}
\affiliation{$^{7}${Collaborative Innovation Center of Extreme Optics, Shanxi University,
Taiyuan, Shanxi 030006, China} }

\begin{abstract}
%the context for your findings 2-3 sentence
%The moiré lattice has recently gained significant attention across solid-state physics, photonics, and cold atom physics,
%as it provides a novel platform for exploring exotic phenomena related to the manipulation of quantum states.
%the problem 1 sentence
%In the realm of neutral cold atoms, while moiré lattices in two-dimensional and three-dimensional systems have been proposed, the simpler one-dimensional moiré physics have yet to be explored.
%here we show... a summarizing sentence
%Here, we propose a scheme that demonstrates moiré effects in a one-dimensional cold atom-cavity coupling system.
%what did we do and what find out, approach and the main result, specific
%This system resembles a generalized open Dicke model that exhibits superradiant phase transitions.
%We uncover a close relationship between the critical point of the phase transition and the one-dimensional moiré parameter.
%Notably,
%we derive the cavity field spectrum,
%which is connected to the dynamical structure factor showcasing the moiré effects.
%We also illustrate a one-dimensional analog of superconductivity governed by the twist angle through the study of atomic diffusion.
%what does it mean, the advance over previous work and implications, specific
%This work provide a new route to test one-dimensional moir\'e effects with cold atoms
%and will enrich the physics of
%moir\'e metrology.
%since the takes place in a one-dimensional moire superlattice, we then term it
%as the moire superradiant phase transition. address its relation with Dicke model.
As a novel platform for exploring exotic quantum phenomena, the moiré
lattice has garnered significant interest in solid-state physics, photonics,
and cold atom physics. While moiré lattices in two- and
three-dimensional systems have been proposed for neutral cold atoms, the
simpler one-dimensional moiré effect  remains largely unexplored. We
present a scheme demonstrating moiré effects in a one-dimensional cold
atom-cavity coupling system, which resembles a generalized open Dicke model
exhibiting superradiant phase transitions. We reveal a strong link between the
phase transition critical point and the one-dimensional moiré
parameter. 
Evidences of the one-dimensional moiré effect are explicitly explored,
including cavity field spectrum,
phase transition dynamics,
and anomalous atomic diffusion.
% Additionally, we derive the cavity field spectrum, connected to the
% dynamical structure factor, and showcase controlled atomic diffusion.
%a one-dimensional analog of superconductivity driven by the twist angle via atomic diffusion.
This work
%paves the way
provides a new route  for testing one-dimensional moiré effects with
cold atoms and open new possibility of
%enriching the physics of
moiré metrology.

\end{abstract}
\maketitle

%\email{lanzhihao7@gmail.com}

%\author{Lu Zhou$^{1}$
%and Weiping Zhang$^{2,5,6}$}

%\section{Introduction}

%let's write the paper
%say that's 2d moire material, we are example of 1d, from this point
%previous correlated insulating phase
\section{introduction} % (fold)
\label{sec:introduction}

% section introduction (end)
Two-dimensional (2D) moir\'{e} lattices engineered by stacking two 2D periodic
layers with a relative twisting angle have emerged as an intriguing new
experimental platform in solid-state physics and optics, in which many exotic
phenomena have been unraveled, including unconventional superconductivity
\cite{rafi2011,Cao2018,tarnopolsky2019}, quantum Hall effect \cite{Dean2013},
non-Abelian gauge potentials \cite{sanjose2012}, and localization
\cite{Wang2020}.
%cold atom, cite some works
%In cold atom community,
As a powerful playground for quantum simulation \cite{bloch2008}, efforts are
pushing forward towards implementing moir\'{e} lattice via trapping neutral
cold atoms in 2D and three-dimensional optical lattices
\cite{ciracPRA2019,lewensteinPRL2020,lewensteinPRB2020,luoPRL2021,Meng2023,wang2024threedimensionalmoirecrystal}%
, which essentially focus on the moiré physics of flat band.
%2d->1d
Noteworthy while one-dimensional (1D) superlattice have already been
implemented in cold atoms to demonstrate localization
\cite{Roati2008,schreiberScience}, the 1D moiré effects in which
moiré parameter would play a vital role in the system quantum
properties remain largely unexplored. Whilst in the counterpart electronic and
optical systems, the research on 1D moiré effects are burgeoning
\cite{dasarmaPRL2021,Talukdar2022,yuan2023,Gonçalves2024,yuanPRL2024,sciadvCavity}.
%In the meanwhile,
%recently a few works have been devoted to one-dimensional (1D) analogs of moir\'{e} lattice \cite{dasarmaPRL2021,Talukdar2022,yuan2023,Gonçalves2024},
%why 1D?
%aimed at demonstrating moir\'{e} physics in the more compact and tractable 1D geometry.
%Inspired by these works,
Question naturally arises on whether one can observe moir\'{e} effects in a 1D
setup with cold atoms.
%moir\'{e} lattice.
%Bearing these ideas in mind,
%Here we propose to study moir\'{e} effects in the system of cold atoms trapped in a cavity-assisted 1D moir\'{e} lattice.

\begin{figure}
[h]\centering
\includegraphics{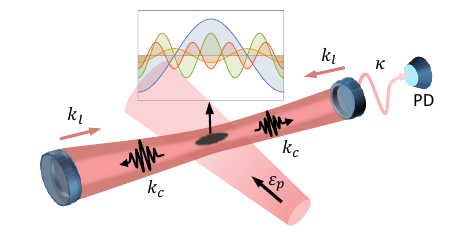}\caption{Schematic diagram for dissipative phase transition in a cavity-assisted moir\'{e} lattice.
Atoms are trapped by a optical lattice potential $V_{l} \cos^{2}(k_{l} x)$ along the cavity axis.
Atoms scatter the side pump laser $\varepsilon_{p}$ into the cavity mode $\cos(k_{c} x+\phi)$.
The inset shows that atoms
 % self-ordering
 are scattered into a few different modes
  in the combined 1D moir\'{e} lattice.
Cavity decay rate is $\kappa$.
}\label{fig:scheme}
\end{figure}

Here we propose a scheme to test such effects, specifically
%superradiance
%at this time comment previous work
in a cold atom-cavity coupling system enabling superradiant phase transition.
%The interplay between
%cavity induced long-range interaction and quantum nature of cold atom gas
%have broadened horizons of studies in cavity quantum electrodynamics \cite{cavityReview1,cavityReview2}.
%Recent years have witnessed new opportunities in cavity quantum electrodynamics brought by trapping cold atoms in an optical resonator \cite{cavityReview1,cavityReview2}.
%coherence are built in the atomic centre-of-mass motion via photon scattering,
%Numerous new results have been unraveled,
%e.g.,
%optical and matter wave nonlinearity \cite{gupta2007,zhou2009},
%exotic quantum phases \cite{larson2008},
%dynamical gauge field \cite{dong2014,lev2019}.
%dicke model, superradiance
%In particular,
%\section{diffusion phase transition in a cavity-assisted moir\'{e} lattice}
%\label{sec:diffusion phase transition} We consider the model depicted in
As sketched in Fig.~\ref{fig:scheme},
%At zero temperature
a cold atom gas composed of $N$ atoms of mass $m$ are trapped
%by a one-dimensional (1D) optical lattice
along the axis of a standing-wave optical resonator in the $x$-direction, and
also illuminated by a laser incident from the cavity side. The driving laser
is detuned far below the atomic transition, thus are scattered by the atoms
into the cavity mode with a pump strength $\varepsilon_{p}$.
%here comment dicke phase transition
The system can be mapped to the well-known Dicke model in quantum optics
\cite{Baumann2010,nagy2010,hemmerichpnas}.
%can be mimiced
Upon driving the pump across a critical value, the interplay between
cavity-mediated global interaction among atoms and cavity decay would bring
the system into a steady state with diverging excitations,
%resemble a generalized Dicke model,
%phase transition takes place
in which photons are collectively scattered into the cavity mode and
%In the meanwhile
the atoms self-organize themselves in the combined net potential of trapping
and emerging intracavity standing-wave potential. A lot of studies on systems
of this type have been reported, both in theory
\cite{nagyPRA2011,Sandner_2015,MASCHLER2007446,cosme2018,zheng2018,yin2020,mivehvar2024,yiweiPRL,boseglasscavity}
and experiment
\cite{esslingerNC2015,klinder2015,Landig2016,landini2018,vaidya2018}.

%figure scheme=================================================

%==============================================================

The paper is organized as follows:
In Section~\ref{sec:moiré_superradiance_and_extended_dicke_model} we introduce the model implementing moiré superradiance,
which can be mapped onto an extended open Dicke model.
The mean-field solution and excitation properties are studied in Section~\ref{sec:mean_field_solution_and_excitations},
with which we unravel the 1D moiré effect and the physics beneath.
In order to verify the moiré effect in a finite quantum system,
we present numerical simulation using truncated Wigner approximation (TWA) in Section~\ref{sec:twa_simulation}.
Section~\ref{sec:cavity_field_spectrum} is devoted to the calculation of cavity field spectrum,
which can be observed in experiment and serve as an moiré effect indicator.
In addition to that,
% the superradiant phase transition,
the moiré effect can be manifested via atomic wavepacket diffusion in the moiré potential,
the anomalous atomic diffusion are explored in Section~\ref{sec:moire diffusion}.
Finally we summarize and outlook in Section~\ref{sec:conclusion}.

%To gain insight into how moir\'{e} physics play a role in the present system,
\section{moiré superradiance and extended Dicke model} % (fold)
\label{sec:moiré_superradiance_and_extended_dicke_model}

% section moiré_superradiance_and_extended_dicke_model (end)
To build the moir\'{e} lattice, we consider an additional one-dimensional (1D)
optical lattice is applied on the atoms,
%In this work
the optical lattice potential $V_{l} \cos^{2}(k_{l} x)$ is chosen
%to be
% incommensurate to
different from
 the intracavity standing-wave mode $\cos(k_{c} x+\phi)$.
Then ratio $r=k_{c} / k_{l}$ defines the moir\'{e} ratio of a combined 1D
bichromatic moir\'{e} lattice \cite{dasarmaPRL2021}.
%say sth on the fibonaci sequence
Specifically we set $r$ as the ratio of two consecutive numbers of the
Fibonacci sequence, i.e., $r = f_{n+1} / f_{n}$. Fibonacci sequence is defined
by the recursion relation $f_{n+1} = f_{n-1} + f_{n}$ with $f_{0} = 0$ and $f_{1} =
1$, in the $n \rightarrow\infty$ limit $r$ approaches the golden ratio $(
\sqrt{5} + 1 ) / 2$. The 1D moir\'{e} lattice is quasiperiodic with a
%periodicity of $M = f_{n+1}$ sites.
unit cell length $L = 2 \pi/ k_{0}$, where $k_{0}$ denotes the elementary wave
number with $k_{l(c)} = f_{n(n+1)} k_{0}$.
%and $M = f_{n+1}$ is
Here we introduce the lowest possible integer numerator of $r$
%, and
%$M$ is termed
as the moir\'{e} parameter $M$ ($= f_{n+1}$), in the following we illustrate
with three cases of $M = 1,3,5$, correspond to $n = 1,3,4$ respectively. For
simplicity we assume the relative phase $\phi= 0$.

%give the master eq here
The dynamics of the joint atom-cavity density operator follows from the master
equation \cite{cavityReview1,cavityReview2}
\begin{equation}
\dot{\rho}=\frac{1}{i\hbar}[H,\rho]+\kappa(2a\rho a^{\dagger}-a^{\dagger}%
a\rho-\rho a^{\dagger}a), \label{eq:master}%
\end{equation}
where
\begin{align}
H  &  =H_{C}+H_{A}+H_{AC},\nonumber\\
H_{C}/\hbar &  =-\delta_{c}a^{\dagger}a,\nonumber\\
H_{A}/\hbar &  =\int dx\Psi^{\dagger}(x)\left[  -\frac{\hbar}{2m}\partial
_{x}^{2}+V_{l}\cos^{2}(k_{l}x)\right]  \Psi(x),\nonumber\\
H_{AC}/\hbar &  =\int dx\Psi^{\dagger}(x)\left[  U_{0}a^{\dagger}a\cos
^{2}(k_{c}x+\phi)\right. \nonumber\\
&  \left.  + \varepsilon_{p}(a+a^{\dagger}) \cos(k_{c}x+\phi)\right]  \Psi(x).
\label{eq:hamiltonian}%
\end{align}
Here the cavity mode is described by an annihilation operator $a$, which
subject to
%pump with a strength $\varepsilon_p$ and
decay with a rate $\kappa$. $\delta_{c}$ is the cavity-pump detuning.
$\Psi(x)$ is the atom bosonic field operator. $U_{0}$ stands for the light
shift per intracavity photon, in the case of frequency redshift $U_{0} < 0$.

%moire physics why do we do that?
%These provide advantage to theoretical treatment, in which
The atom field can be effectively expanded with a finite number of modes
\begin{equation}
\Psi(x) = \sqrt{ \frac{1}{L} } c_{0} + \sum_{n=1}^{n_{c}} \sqrt{ \frac{2}{L} }
\cos(n k_{0} x) c_{n}, \label{eq:modes}%
\end{equation}
where the $c_{0(n)}$ are bosonic annihilation operator. We have precluded the
odd parity (sine) modes by considering the parity symmetry for bosons
initially in a Bose-Einstein condensate. Note that collisional atom-atom
interactions can be tuned small and they only play a role in shifting the
single particle dispersion \cite{Baumann2010}, so we neglect collisional
interactions here without affecting the main physics. A cutoff $n_{c}$ is
introduced due to that high energy modes are less likely to be excited. Insert
(\ref{eq:modes}) into (\ref{eq:hamiltonian}), by evaluating the integrals
within one unit cell we can obtain the Hamiltonian in a reduced subspace as
\begin{align}
H / \hbar &  = -\delta_{c}a^{\dagger}a + \omega_{k} c^{\dagger}\mathcal{K} c +
\frac{V_{l}}{4} c^{\dagger}\mathcal{M}_{2 f_{n}} c\nonumber\\
+ \frac{U_{0}}{4}  &  a^{\dagger}a c^{\dagger}( 2 \mathds{1} + \mathcal{M}_{2
f_{n+1}}) c + \frac{\varepsilon_{p}}{2} (a + a^{\dagger}) c^{\dagger
}\mathcal{M}_{f_{n+1}} c, \label{eq:hami1}%
\end{align}
where $c = (c_{0}, c_{1}, \dots, c_{n_{c}})^{T}$, the recoil frequency
$\omega_{k} = \hbar k_{0}^{2} / 2 m$, $\mathds{1}$ is the identity matrix, the
matrices $\mathcal{K}$, $\mathcal{M}_{f_{n}}$ are given in 
Appendix~\ref{app_matrix}.
% the supplemental
% material.
%\cite{sm}.
Since in real experiment cavity mode is usually fixed, then in the following
we scale the energy with $\omega_{r} = \hbar k_{c}^{2} / 2 m = M^{2}
\omega_{k}$. That is to say, moir\'{e} parameter is manipulated by $k_{l}$.
%say sth on extended dicke model
Hamiltonian (\ref{eq:hami1}) typically represents an extended Dicke model, in
which the quantum light field is effectively coupled to multilevel
transitions.
%of many atoms.
In addition to scattering within the cavity, the externally applied lattice
allows for momentum transfers among atoms, giving rise to moir\'{e} effects in
atom self-ordering and superradiant phase transition as will be illustrated
below.
%As an important feature the Hamiltonian allows for momentum
%transfers of in addition to from scattering within the cavity. This enables
%the system to self-order and has also important consequences for the cooling behavior.
%\subsection{mean-field solution and excitations}
%\label{sub:mean-field solution and excitations}

\section{mean field solution and excitations} % (fold)
\label{sec:mean_field_solution_and_excitations}

\begin{figure}
[htb]\centering
\includegraphics[width=3 in]{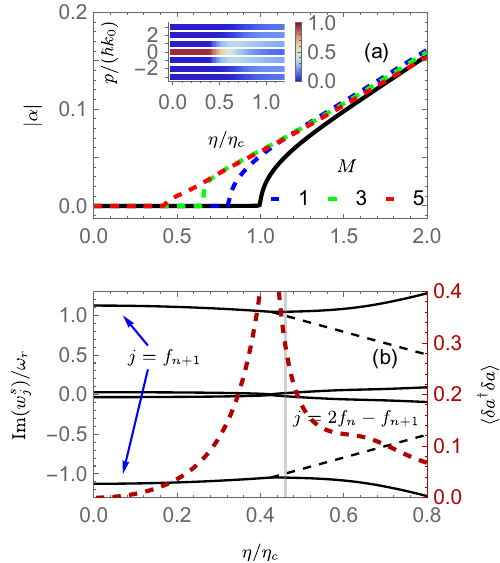} \caption{
(a) Scaled mean-field value $| \alpha |$ versus scaled pumping strength $\eta / \eta_c$ for different $M$.
The black solid line is the case without the external optical lattice.
The inset present steady state momentum distribution for $M = 5$.
(b) The eigenvalues imaginary parts of atom polariton excitations
 versus $\eta$ are shown for 
 two major scattering channels at $M = 5$,
 in which the absolute values of the pair $j = 2 f_n - f_{n+1}$ is much smaller than those of $j = f_{n+1}$,
 thus determining the phase transition point.
 The pair $j = f_{n+1}$ would converge to $0$ at around $\eta = \eta_c$ without the occurrence of phase transition,
 as indicated by the black-dashed lines.
The vertical grey line indicate the analytical phase transition point.
Red-dashed line showcase incoherent cavity excitation.
The steady states and excitations are calculated for
$V_l = -1$,
$N U_0 = -80$,
$\Delta_c = -100$ and
$\kappa = 20$.
}\label{fig:dissipation}
\end{figure}

% section mean_field_solution_and_excitations (end)
% \emph{Mean-field solution and excitations.---} 
In the $N \rightarrow\infty$
mean-field limit, the operators $a$, $c_{ 0(n) }$ split into
\begin{equation}
a = \sqrt{N} \alpha+ \delta a,~ c_{ 0(n) } = e^{ - i \mu t } \left[  \sqrt{N}
\psi_{ 0(n) } + \delta c_{ 0(n) } \right]  , \label{eq:mean-field}%
\end{equation}
where $\alpha$, $\psi_{ 0(n) }$ are scaled steady state expectations with
$\delta a$, $\delta c_{ 0(n) }$ characterize the corresponding quantum
fluctuations, $\mu$ is the chemical potential. By solving stationary equations
containing $\alpha$ and $\psi_{ 0(n) }$ 
(see Appendix~\ref{app_mean field})
% \cite{sm}
, the cavity field amplitude
$| \alpha|$ are plotted in Fig.~\ref{fig:dissipation}(a) versus the scaled
pumping strength $\eta= \sqrt{N} \varepsilon_{p}$.

Without the external optical lattice potential, the present system exactly
resemble the open Dicke model \cite{nagyPRA2011}. In the thermodynamical limit
the mean-field solutions would predict a critical pumping strength $\eta_{c} =
\sqrt{-( \Delta_{c}^{2} + \kappa^{2} ) \omega_{r} / 2 \Delta_{c}}$ with
effective detuning $\Delta_{c} = \delta_{c} - N U_{0}/2$, which separates the
normal phase $\left\{  \alpha= 0, \psi_{0} = (1, 0, 0, \dots)^{T} \right\}  $
from the superradiant phase with a finite $\alpha$.
%$|\alpha| > 0$.
In Fig.~\ref{fig:dissipation}(a) this case is indicated by a black solid line.
With the increase of moir\'{e} parameter $M$, one can see that the
corresponding critical pumping gradually decreases. Upon the onset of
superradiant phase transition, in the steady state atoms disperse from the
homogeneous state into the modulated states in which atoms occupy modes of discrete momentum $p$, as illustrated in the inset of
Fig.~\ref{fig:dissipation}(a) for $M = 5$.

%It would be insightful to look into the excitation properties.
The relation between the critical pumping strength and the moir\'{e} parameter
$M$ can be understood via a study on the excitation properties presented in
Appendix~\ref{app_excite}
% the End Matter
, in which we derive the atom polariton excitation frequencies $w_j^s$,
with whose lowest imaginary value approach $0$ we determine the phase transition critical point,
as shown in Fig.~\ref{fig:dissipation}(b).
In addition to that
% the usual scattering
% channel involving 
atoms in the homogeneous state scatter the pump field 
% is scattered
 into the cavity mode and give rise to atom polariton excitation $w_{j=f_{n+1}}^s$,
% add one more curve in the figure,
% compare the two frequency,
% this problem can be solved.
% of
% $\left\{  \delta a, \delta c_{0(n)} \right\} $.
% A phase transition occurs if the lowest
%\cite{sm}:
%A simple explanation on this effect would be that
%decreasing the lattice constant means that the atoms are more tightly bounded and require more energy to move,
%leading to an increased band gap,
%while
the moiré lattice provides an additional scattering channel hiring atoms in
the $\cos(2 k_{l} x)$ mode,
 % are scattered by the pump field into the cavity
% photon, 
resulting in atomic polariton excitations with eigenfrequency
$w_{j=2f_{n}-f_{n+1}}^s$ whose absolute imaginary value is much smaller.
% less than that of the usual channel with $j=f_{n+1}$.
%with the moiré ratio we chosen here.
Increasing $M$
%means an
% effectively equals enlargement of the lattice constant, which 
effectively decrease
the energy gap and thus facilitating superradiant phase transition. In
Fig.~\ref{fig:dissipation}(b) we mark the analytic critical value for $M=5$ as
the vertical line, which approximately coincide with the point at which the
imaginary parts of the lowest atom polariton excitation frequency ($j = 2 f_n - f_{n+1}$) reach zero. Incoherent cavity field excitation $\left\langle \delta
a^{\dagger}\delta a\right\rangle $ (red-dashed line) also becomes divergent
upon the onset of phase transition.

% ==fig mean-field=============
% =============================

\section{twa simulation} % (fold)
\label{sec:twa_simulation}

% section twa_simulation (end)
% \emph{Truncated Wigner approximation (TWA) simulation.---} 
In order to verify
whether the phenomena predicted above in the thermodynamical limit can really
take place in a finite quantum system, here we simulate the dynamics in the
open system depicted by the master equation (\ref{eq:master}). Due to the in
principle unconstrained Hilbert space dimension of the cavity field in a
generalized Dicke model and its global coupling to all the atomic modes, a
full quantum simulation hiring Monte-Carlo wavefunction (quantum jump)
algorithm is usually limited to small system size
\cite{Sandner_2015,kollathPRR2020}. We adopt TWA to study the dynamics with the details given in the supplemental material
\cite{sm}. The TWA method
\cite{Alice_Sinatra_2002,GardinerReview,POLKOVNIKOV20101790,AMReyPRX2015} have
found qualitative agreements with experimental results in a series of
atom-cavity setup of time crystal experiments
\cite{Cosme2021a,Cosme2021b,Cosme2023}.
%find out the order parameter which can probe the moire superradiant phase transitionsss
%==================

%fig-twa present the twa simulation results===========
\begin{figure}
[htb]\centering
\includegraphics[width=3 in]{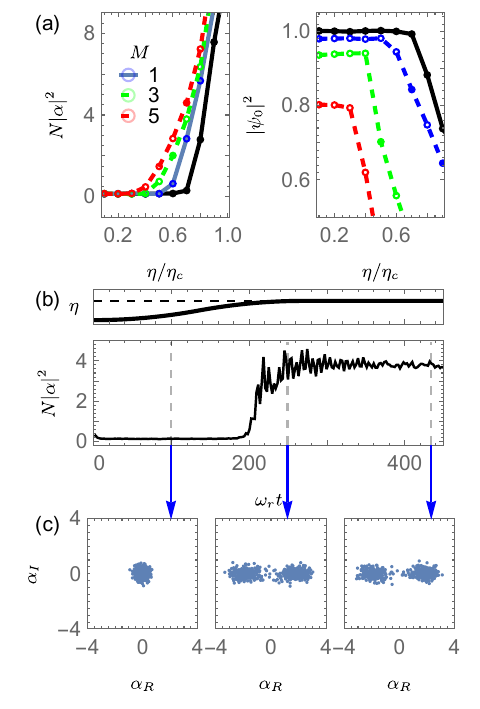} \caption{
TWA simulation results with $N=1000$ obtained from sample of $500$ trajectory runs.
(a) Steady state population in cavity mode (left panel) and homogeneous atomic mode $\psi_0$ (right panel) versus pumping strength $\eta$.
The black solid line is the case without the external optical lattice.
(b) Cavity field population dynamics at $M = 3$ along with $\eta$ ramps to the value of $0.6 \eta_c$.
(c) Wigner distributions of cavity light field at
the instant marked by the vertical grey dashed line in (b).
The parameters are the same as those in Fig.~\ref{fig:dissipation},
noticeably that TWA however predicts the superradiant phase transition to occur at lower $\eta$.
}\label{fig:twa}
\end{figure}

%===================================

%figure cavity spectrum===========
%\begin{widetext}
\begin{figure*}[htb]
\centering
\includegraphics[width=6 in]{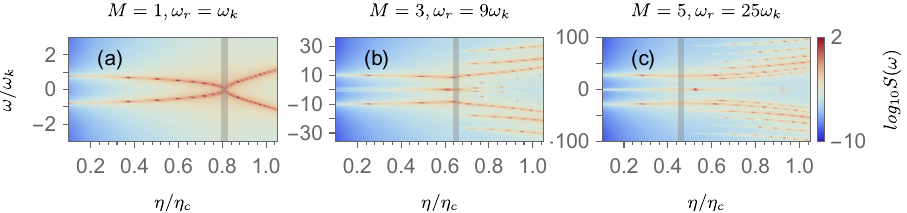}
\caption{
The logarithm of the cavity field spectrum
$S(\omega)$ as a function of $\omega$ and relative pumping $\eta$.
% in which the coherent $\omega = 0$ part have been omitted.
From left to right:
(a) $M = 1$,
(b) $M = 3$,
(c) $M = 5$.
The vertical gray bar indicate the critical pumping strength at which phase transition takes place for each cases.
Below threshold two pairs of sideband peaks are visible,
corresponding to frequencies of two types of quasi-particle excitations as specified in the main text.
% The energy of the lower quasi-particle excitations vanishes towards the critical point.
Upon the onset of superradiant phase transition,
more peaks appear and 
they are intimately related to 
% larger 
the moiré parameter $M$.
 % would give rise to more excitation modes.
 }%
\label{fig:sw}%
\end{figure*}
% \end{widetext}
% =================================

%i should address the symmetry problem,
%what's the link between symmetry and phase transition problem?
%say the results
As shown in Fig.~\ref{fig:twa}(b), in every run of TWA simulation we propagate
an initial state of $N = 1000$ atoms in the homogeneous mode $\psi_{0}$ while
the other atomic modes and the cavity field are left empty,
%during which the
%cavity pumping is ramped
by ramping the cavity pumping up to a desired value $\eta$ and holding it
there until a steady state is reached. Compared with mean-field results, the
TWA calculation indicate that onset of superradiant phase transition will
apparently takes place at a smaller critical pumping $\eta$, as illustrated in
Fig.~\ref{fig:twa}(a). This suggests that quantum fluctuations will lower the
critical value for the phase transition to take place. Apart from that, the
dependence of superradiant phase transition on the moir\'{e} parameter is
verified.
%Especially
In the right panel of Fig.~\ref{fig:twa}(a), for $\eta= 0$
%the well-known spectrum of is rendered.
steady state population in homogeneous mode $\psi_{0}$ becomes smaller with
increasing $M$.
%since the
This behavior is because the external optical lattice scatter the $\psi_{0}$
atoms into $\psi_{2 k_{l}}$, whose scaled kinetic energy $4 k_{l}^{2} /
k_{c}^{2} = 4/r^{2}$ decreases when $M$ becomes larger.
%below phase transition threshold the steady state
%is less than unity
%at steady state
%with $M$ increases.
%as indicated in the figure at $\eta = 0$.
The larger seed population of $\psi_{2 k_{l}}$ in turn triggers dynamical
instability of the normal phase at a smaller pumping strength $\eta$.

%can i understand due to the seeds becomes larger,
%it will facilitate phase transition.
%say the photon statistics
Although the presented TWA results already showcase the properties of phase
transition, it is argued that finite system size will smoothen the abrupt
change upon phase transitions in the thermodynamical limit \cite{Sandner_2015}%
, rendering that neither $| \alpha|^{2}$ nor $| \psi_{0} |^{2}$ a good phase
transition indicator. Phase transitions are accompanied with spontaneous
symmetry breaking. In the absence of the external optical lattice (Dicke model), with the
occurrence of superradiance and the formation of intracavity optical lattice,
atoms can spontaneously choose even or odd lattice sites to reside, which
entangle with cavity field of opposite amplitudes and result in a macroscopic
superposition state of $\vert\text{even} \rangle\otimes\vert\alpha\rangle+
\vert\text{odd} \rangle\otimes\vert-\alpha\rangle$
\cite{DomokosPRL2002,MASCHLER2007446}. Similar processes take place for the
present moir\'{e} lattice. For the dynamics of above threshold pumping in the
$M = 3$ case studied in Fig.~\ref{fig:twa}(b), we project the cavity field
sample of $500$ trajectories at three different times onto Fig.~\ref{fig:twa}%
(c), which unravel its Wigner distribution. The cavity field evolves from
vacuum noise (left panel), then being stretched and splitted (middle panel),
and finally forms a Schr\"{o}dinger cat state with opposite amplitudes (right
panel). The associated atom steady states would be that atoms located at a few
different sites of the combined moir\'{e} lattice. Due to the inherent
symmetry of steady state, $\langle a \rangle$ or equivalently $\langle
c^{\dagger}\mathcal{M}_{f_{n+1}}c \rangle$ would give the value $0$. To break
the symmetry, one can project the system state with respect to one cavity
state maximizing the Wigner distribution and resolve the corresponding order
parameters \cite{Sandner_2015,OstermanPRL2020}.

%\subsection{cavity field spectrum}
%\label{sub:cavity spectrum}
\section{cavity field spectrum} % (fold)
\label{sec:cavity_field_spectrum}

% section cavity_field_spectrum (end)
% \emph{Cavity field spectrum.---}
%say that cavity spectrum can be detected
The onset of phase transition can be indicated by the dynamic structure
factor, which is the Fourier transform of the intracavity field correlator
$S(\omega) \equiv\mathcal{F}\left(  \left\langle a^{\dagger}\left(
t_{s}+t\right)  a\left(  t_{s}\right)  \right\rangle \right)  $ ($t_{s}$ is
the time for the system to reach steady state)  and thus can be probed in
experiment \cite{esslingerNC2015,agarwal2006}. We calculate $S(\omega)$ in 
Appendix~\ref{app_spectra}
% the supplemental material
% \cite{sm} 
with the results presented in Fig.~\ref{fig:sw}.

To map the dynamical structure factor, in the spectra plotted we have dropped
the coherent part of $\omega= 0$ \cite{esslingerNC2015}, which would become
prominent upon the onset of superradiant phase transition as photons are
collectively scattered into the cavity mode. The spectra are almost symmetric
with respect to $\omega= 0$ and their peaks come in pairs.
%however for a pair of the red-shifted and the blue-shifted sideband,
%their spectra weight are not equal.
%what does the peak mean? give some idea here
Peaks of cavity spectrum actually reflect frequencies of atomic polariton mode
excitation consisting of photonic and atomic parts \cite{sm}. In the regime
far below the threshold, the spectra peaks are located at $\omega\simeq
\pm\omega_{r}$ and $\pm\omega_{k} = \pm\hbar(2k_{l} - k_{c})^{2} / 2m$,
reflecting that the cavity photons are scattered from the atoms in the
homogeneous mode and those in the mode of $\cos( 2 k_{l} x )$, respectively.
With the increase of pumping strength, the peaks in both pairs gradually move
toward each other, the $\pm\omega_{k}$ peaks even merge at the critical point.
This behavior is due to that the photonic components are becoming larger in the
atomic polariton modes.
 % and 
 Upon the phase transition, the lowest
atomic polariton excitation frequency approaches zero.
 These behaviors are consistent with those of atomic polariton excitations presented in
Fig.~\ref{fig:dissipation}(b). Beyond the critical point, more pairs of peaks
appear with the increase of the moir\'{e} parameter $M$, and the peaks of a
pair tend to repel from each other when the pumping is continuously enhanced.
The phenomena root in the fact that more atomic polariton modes are excited
and their energies increase as well.
%which was also indicated in
%Fig.~\ref{fig:dissipation}(b).
The intracavity field spectra 
% (dynamical structure factor)
 predicted here
%display clear
can serve as evidence of moir\'{e} effects on the superradiant phase
transition.

\begin{figure}
[htb]\centering
\includegraphics[width=3 in]{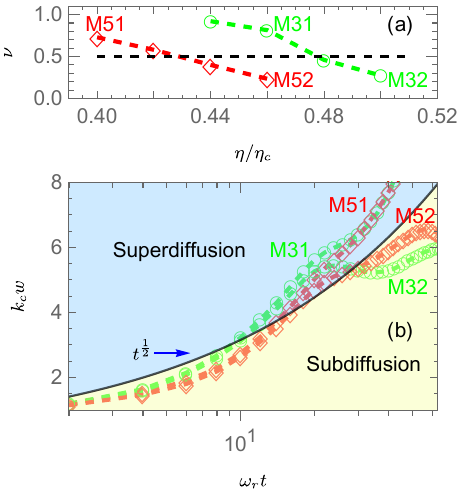} \caption{
(a) Time scaling $\nu$ of atomic diffusion versus pumping strength $\eta$ for $M = 3$ (green-dashed line) and $M = 5$ (red-dashed line).
The horizontal black-dashed line of $\nu = 1/2$ indicates the critical scaling.
(b) Atomic diffusion dynamics simulated for
M31 ($M = 3, \eta = 0.46 \eta_c$), M32 ($M = 3, \eta = 0.5 \eta_c$), M51 ($M = 5, \eta = 0.4 \eta_c$) and M52 ($M = 5, \eta = 0.46 \eta_c$),
which were also indicated in (a).
The black line indicates the normal diffusion of $\nu = 1/2$ that separates superdiffusion ($\nu > 1/2$) and subdiffusion ($\nu < 1/2$) regions.
Atoms are initially prepared in a Gaussian wavepacket with width $w = 1/k_c$.
The other parameters are the same as before.
}
%be careful on the
%scale, this figure give the phase transition, mode occupation and excitation
%number
\label{fig:diff}
\end{figure}

%==================================

\section{anomalous diffusion dynamics}
\label{sec:moire diffusion}
%\emph{DPT and the QFI.---}
%diffusion, let's finish it!
%finally we arrived at this, it's close to an end
%the problem is program
% \emph{Atomic diffusion.---} 
Considering the long time for the system to reach
steady state versus limited lifetime of cold atoms, on the atomic part it
would be more practicable to observe moiré effects through their
diffusion. The dependence of superradiant phase transition on the moir\'{e}
parameter provides an extra control knob on atom diffusion.
%opportunity to display a phenomenon analogue to that
%superconductivity controlled by the twist angle in 2D moir\'{e} systems.
This can be implemented by loading atoms into the cavity and observing
%its
their transport along the moir\'{e} lattice. We study the atom diffusion by
taking the initial atomic wavefunction to be a Gaussian wavepacket
%(with even parity)
%localized
%in the center of a unit cell
with width $w$, while the cavity is in vacuum. Then the pump is tuned on and
the wavepacket width is estimated with $w=\sqrt{\left\langle \left(  \Delta
x\right)  ^{2}\right\rangle }$ (see
Appendix~\ref{app_diff} 
% supplemental material \cite{sm}
 for details).

For atomic wavepacket,
the time evolution of its width $w(t)$ can be parametrized as $w(t) \sim t^{\nu}$
\cite{lancher2009,hiramoto1988,zhou2013}. $\nu=1$ correspond to ballistic
expansion. 
% Under the combined effects of moir\'e lattice and cavity field
% quantum fluctuation, atoms would exhibit anomalous diffusion. 
We extract the
time scaling $\nu$ from the atomic diffusion dynamics and present the results
in Fig.~\ref{fig:diff}(a). 
% As expected, 
The time scaling $\nu$ gradually decreases when
pumping $\eta$ becomes stronger.
This behavior is in contrast with the  
% predicted with 
mean-field theory results,
which predict an abrupt jump of $\nu$ from the value of $1$ (ballistic expansion) to approximately $0$ (wavepacket localization) upon the onset of superradiant phase transition,
 signaling
anomalous diffusion.
Via mapping to the mean-field critical points,
%atomic diffusion
% the system transfers from the
we identify two
 regimes of anomalous diffusion:
 superdiffusion ($\nu > \frac{1}{2}$) versus
subdiffusion ($\nu < \frac{1}{2}$). 
% For $M = 3$ and $5$,
The boundary separating these two regimes are determined by the intersection between the horizontal line of $\nu = \frac{1}{2}$ and the corresponding time scaling curves,
as shown in Fig.~\ref{fig:diff}(a).
% As expected,
% The boundary of the two regimes is identified with $\nu=
% \frac{1}{2}$. The superdiffusion to subdiffusion transfer is accompanied with
% the onset of superradiant phase transition and buildup of moiré
% lattice.
%Then
%it is self-evident
%one can understand
%that atomic diffusion would be related to the moiré parameter.
% This observation suggests a relation between atom diffusion and moiré
% pattern. As shown in Fig.~\ref{fig:diff}(a), the scaling behavior indicate
The superdiffusion to subdiffusion transition takes place at a smaller $\eta$
for the $M = 5$ case (red-dashed line) as compared with that of $M = 3$
(green-dashed line). 

% This is consistent with our above analysis on the
% relation between moiré parameter and superradiant phase transition.

More specifically, 
we exemplify the diffusion dynamics in Fig.~\ref{fig:diff}(b).
% In the case of weak and strong 
With the increase of
pumping $\eta$,
corresponding to M51 ($M = 5$, $\eta = 0.4 \eta_c$) $\rightarrow$ M52 ($M = 5$, $\eta = 0.46 \eta_c$),
and M31 ($M = 3$, $\eta = 0.46 \eta_c$) $\rightarrow$ M32 ($M = 3$, $\eta = 0.5 \eta_c$),
% respectively,
one can observe that the diffusion 
property varies from
% fall well within the regime of 
superdiffusion to subdiffusion.
Apart from that,
% However 
moiré effect also matters in atomic diffusion.
For M31 and M52 which
% , as indicated in
% Fig.~\ref{fig:diff}(a), they 
subject to identical pumping of $\eta= 0.46
\eta_{c}$ however different $M$,
% we examine two cases of 
%for M31 and M52 in which the dynamics are simulated under the very same $\eta$ however different $M$,
M31 displays superdiffusion while M52 behaves subdiffusion, as demonstrated in
Fig.~\ref{fig:diff}(b). 
The anomalous diffusion behavior is amount to the combined effects of moir\'e lattice and cavity field
quantum fluctuation.

\section{discussion and outlook}
\label{sec:conclusion}
% \emph{Discussion and outlook.---}
%summerize the result
We have studied moir\'{e} effects on superradiant phase transition in a cold
atom-cavity coupling system. This system resembles a generalized open Dicke
model.
%The intimate relation between
The prominent role of moir\'{e} parameter in superradiant phase transition is
explicitly explored.
%experiment consideration
%Our scheme can be
%readily implemented with the current state-of-art experimental techniques of
Implementing an optical lattice for $^{87}$Rb Bose-Einstein condensate  inside
a high-finesse optical cavity is feasible with existing technology
\cite{klinder2015,Landig2016,vaidya2018}.
%It provide a new route to test 1D moir\'e effects with cold atoms.

To observe the moiré effects, one can either utilize the optical means
of cavity transmission spectrum or monitor atom diffusion.
%superfluid fraction
%We demonstrated the
%controlled atomic diffusion via moir\'{e} parameter.
We have demonstrated that atom diffusion is controlled by the moiré
parameter, and provides an ideal experimental observable to reveal the effects
produced by the moiré pattern.
%1D analogue of superconductivity controlled by the twist angle
%via studying the atomic diffusion.
In future work it will be interesting
%To probe
to look into the combined effects of moir\'{e} geometry and quantum
fluctuations
%the effect of quantum geometry
on superfluidity,
%it will be necessary
in which one will need to estimate physical quantities such as superfluid
fraction (weight)
\cite{castin2011,Sidorenkov2013,Ho2010,john2011,cooper2011,esslinger2013,Biagioni2024}%
. Besides that,
%fermion statistics
%It will also be interesting to
%consider
one can also expect to observe  moir\'{e} effects in fermionic superradiance
\cite{keeling2014,piazza2014,chen2014,zhang2021,pan2022,wu2023superradianttransitionfermionicquasicrystal}
and many body localization \cite{husePRB,jiePRA,caiPRB,jakubPRB}.
%take Fermi statistics into account
%critical metrology
As criticality can serve as a valuable source for quantum metrology
\cite{zanardi2008,macieszczak2016,garbe2020,chu2021,Ding2022,ilias2022,Garbe_2022,Gietka2022understanding,Aybar2022criticalquantum,guan2021,zhou2023,wang2024}
and intimate relation between moir\'{e} parameter and superradiant phase
transition have been revealed here,  enhanced estimation on the moir\'{e}
parameter would
%also worth further research.
enrich the physics of moir\'e metrology
\cite{kafri1990physics,Post2008,Halbertal2021}.

\begin{acknowledgments}
We thank Han Pu for helpful discussions. This work is supported by the
Innovation Program for Quantum Science and Technology (2021ZD0303200);
National Key Research and Development Program of China (Grant No.
2016YFA0302001), the National Natural Science Foundation of China (Grant Nos.
12074120,
%11374003,
%11654005,
12374328,  12234014, 12005049, 11935012), the Shanghai Municipal Science and
Technology Major Project (Grant No. 2019SHZDZX01), Innovation Program of
Shanghai Municipal Education Commission (Grant No. 202101070008E00099),
Shanghai Science and Technology Innovation Project (No. 24LZ1400600), and the
Fundamental Research Funds for the Central Universities. A.C. acknowledges
funding from the Spanish Ministry of Science and Innovation
MCIN/AEI/10.13039/501100011033 (project MAPS PID2023-149988NB-C21), the EU
QuantERA project DYNAMITE (funded by MICN/AEI/ 10.13039/501100011033 and by
the European Union NextGenerationEU/PRTR PCI2022-132919 (Grant No.
101017733)), and the Generalitat de Catalunya (AGAUR SGR 2021- SGR-00138)).
W.Z. acknowledges additional support from the Shanghai Talent Program.
%National Natural Science
%Foundation of China (Grants No. , No. 11774093) and the National Key
%Research and Development Program of China (Grant No. 2016YFA0302001).
L.Z. acknowledges additional support from China Scholarship Council.
%the Natural Science Foundation of Shanghai (Grant No. 20ZR1418500).

\end{acknowledgments}

\appendix

\section{Matrix definition}

\label{app_matrix}

The matrices $\mathcal{K}$, $\mathcal{M}_{f_{n}}$ denote the kinetic energy
term and the terms propotional to $\cos (f_{n}k_{0}x)$ in the Hamiltonian,
which are defined as%
\begin{align}
\mathcal{K}&=\left( 
\begin{array}{ccccccc}
0^{2} &  &  &  &  &  &  \\ 
& 1^{2} &  &  &  &  &  \\ 
&  & 2^{2} &  &  &  &  \\ 
&  &  & \ddots  &  &  &  \\ 
&  &  &  & \ddots  &  &  \\ 
&  &  &  &  & (n_{c}-1)^{2} &  \\ 
&  &  &  &  &  & n_{c}^{2}%
\end{array}%
\right) ,
\nonumber
\\
\mathcal{M}_{f_{n}}&=\left( 
\begin{array}{ccccccc}
&  &  &  & \sqrt{2} &  &  \\ 
&  &  & 1 &  & 1 &  \\ 
&  & a &  &  &  & \ddots  \\ 
& 1 &  &  &  &  &  \\ 
\sqrt{2} &  &  &  &  &  &  \\ 
& 1 &  &  &  &  &  \\ 
&  & \ddots  &  &  &  & 
\end{array}%
\right) .  \label{eq:matrix}
\end{align}%
In $\mathcal{M}_{f_{n}}$ modes $\cos (j(k)k_{0}x)$ with $j\pm k=f_{n}$ are
coupled, giving a nonzero element. % \begin{equation}
% \begin{pNiceMatrix}[first-row]
%          &   &        &   & (\text{row } f_n) & & \\
%          &   &        &   & \sqrt{2} &   &        \\
%          &   &        & 1 &          & 1 &        \\
%          &   & \adots &   &          &   & \ddots \\
%          & 1 &        &   &          &   &        \\
% \sqrt{2} &   &        &   &          &   &        \\
%          & 1 &        &   &          &   &        \\
%          &   & \ddots &   &          &   &        
%         % 0 & 0 & 0 & 0 & 0 & 0 & 0 & 1
% \end{pNiceMatrix}
% \end{equation}

\section{Solve stationary equation}
\label{app_mean field}

The Heisenberg-Langevin equations for the field amplitude $a$, $c_{0(n)}$
read 
\begin{align}
\dot{a} & =\left[ i\left( \delta_{c}-\frac{U_{0}}{4}c^{\dagger }(2\mathds{1}+%
\mathcal{M}_{2f_{n+1}})c\right) -\kappa\right] a
\nonumber\\
&-i\frac {\varepsilon_{p}}{2}%
c^{\dagger}\mathcal{M}_{f_{n+1}}c+a_{in},  \notag \\
\dot{c} & =-i\left[ \omega_{k}\mathcal{K}+\frac{V_{l}}{4}\mathcal{M}%
_{2f_{n}}+\frac{U_{0}}{4}a^{\dagger}a(2\mathds{1}+\mathcal{M}_{2f_{n+1}}) \right.
\nonumber\\
&\left. +%
\frac{\varepsilon_{p}}{2}(a+a^{\dagger})\mathcal{M}_{f_{n+1}}\right] c, 
\label{eq:Heisenberg-Langevin}
\end{align}
where the Gaussian noise operator $a_{in}$ has zero mean $\langle
a_{in}(t)\rangle=0$ and nonvanishing correlation $\langle
a_{in}(t)a_{in}^{\dagger}(t^{\prime})\rangle=\kappa\delta(t-t^{\prime})$.
Inserting $a=\sqrt{N}\alpha+\delta a,~c_{0(n)}=e^{-i\mu t}\left[ \sqrt{N}%
\psi _{0(n)}+\delta c_{0(n)}\right] $ into Eqs. (\ref{eq:Heisenberg-Langevin}%
), the corresponding mean-field stationary equations can be derived as 
\begin{equation}
\lbrack\Delta(\psi)-i\kappa]\alpha+\frac{\eta}{2}\psi^{T}\mathcal{M}%
_{f_{n+1}}\psi=0,~\mathcal{C}(\alpha)\psi=\mu\psi,   \label{eq:stationary}
\end{equation}
where $\psi=(\psi_{0},\psi_{1},\dots,\psi_{n_{c}})^{T}$, $u=\frac{NU_{0}}{4}$%
, $\Delta_{c}=\delta_{c}-2u$, $\eta=\sqrt{N}\varepsilon_{p}$ 
%$\alpha = \alpha_R + i \alpha_I$,
and 
\begin{align}
\Delta(\psi)&=-\Delta_{c}+u\psi^{T}\mathcal{M}_{2f_{n+1}}\psi,
\nonumber\\
\mathcal{C}%
(\alpha)&=\omega_{k}\mathcal{K}+\frac{V_{l}}{4}\mathcal{M}_{2f_{n}}+u|%
\alpha|^{2}(2\mathds{1}+\mathcal{M}_{2f_{n+1}})
\nonumber\\
&+\eta\operatorname{Re}(\alpha)%
\mathcal{M}_{f_{n+1}}.   \label{eq:s-scale parameter}
\end{align}
Eqs. (\ref{eq:stationary}) can be solved in a self-consistent manner. Start
from an initial guess $\alpha$, diagonalize $\mathcal{C}(\alpha)$ to find
out the $\psi$ corresponding to minimum $\mu$. Use this $\psi$ to derive a
new $\alpha$, then repeat the above process until convergence is reached.

\section{Excitations}

\label{app_excite}

Hiring linear pertubation theory, use the steady-state solutions $\left\{
\alpha,\psi\right\} $ obtained from Section~\ref{app_mean field}, we first
perform an orthogonal tranformation%
\begin{equation}
\delta b_{j}=\mathcal{O}^{T}\delta c_{j},   \label{eq:s-transform}
\end{equation}
with $\delta b\left( c\right) _{j}=\left( \delta b\left( c\right)
_{0},\delta b\left( c\right) _{1},\dots,\delta b\left( c\right)
_{n_{c}}\right) ^{T}$, and $\mathcal{O}$ is a matrix whose $j$-th column
contains the $j$-th eigenvector of $\mathcal{C}(\alpha)$\ with the
corresponding eigenvalue $w_{j}$. The equations for excitations $\left\{
\delta a,\delta b_{j}\right\} $ then read 
\begin{align}
\frac{d}{dt}\delta a & =\left[ -i\Delta\left( \psi\right) -\kappa\right]
\delta a-i\sum_{j}\left( \frac{\eta}{2}g_{j}^{1}+u\alpha g_{j}^{2}\right)
\nonumber\\
&\times
\left( \delta b_{j}+\delta b_{j}^{\dagger}\right) +a_{in},  \notag \\
\frac{d}{dt}\delta b_{j} & =-i\left( w_{j}-\mu\right) \delta b_{j}-i\left[ 
\frac{\eta}{2}g_{j}^{1}\left( \delta a+\delta a^{\dagger }\right)\right.
\nonumber\\
&\left.+ug_{j}^{2}\left( \alpha^{\ast}\delta a+\alpha\delta a^{\dagger }\right) %
\right] ,   \label{eq:s-excitation}
\end{align}
where $g_{j}^{1}=\mathcal{O}^{T}\mathcal{M}_{f_{n+1}}\psi$ and $g_{j}^{2}=%
\mathcal{O}^{T}(2\mathds{1}+\mathcal{M}_{2f_{n+1}})\psi$. Since
$w_0 = \mu$,
rendering 
the
quadrature $\frac{1}{2}\left( \delta b_{0}+\delta b_{0}^{\dagger}\right) $
a constant of motion,
 % and the condensate phase fluctuation $\frac{1}{2i%
% \sqrt{N}}\left( \delta b_{0}-\delta b_{0}^{\dagger}\right) $ would not take
% significant effect in the thermodynamic limit, 
we then neglect the
fluctuation $\delta b_{0}$, deduce the Langevin equations for the operator
vector $\mathcal{V}=\left\{ \delta a,\delta a^{\dagger},\delta b_{1},\delta
b_{1}^{\dagger},\dots,\delta b_{n_{c}},\delta b_{n_{c}}^{\dagger}\right\}
^{T}$ as%
\begin{equation}
\frac{d}{dt}\mathcal{V}=\mathcal{M}_{e}\mathcal{V}+\xi, 
\label{eq:s-langevin}
\end{equation}
with the elementary excitation matrix%
\begin{widetext}

\begin{equation}
\mathcal{M}_{e}=\left( 
\begin{array}{ccccc}
-i\Delta-\kappa & 0 & -i\left( \frac{\eta}{2}g_{1}^{1}+u\alpha
g_{1}^{2}\right) & -i\left( \frac{\eta}{2}g_{1}^{1}+u\alpha g_{1}^{2}\right)
& \cdots \\ 
0 & i\Delta-\kappa & i\left( \frac{\eta}{2}g_{1}^{1}+u\alpha^{\ast}g_{1}^{2}%
\right) & i\left( \frac{\eta}{2}g_{1}^{1}+u\alpha^{\ast}g_{1}^{2}\right) & 
\cdots \\ 
-i\left( \frac{\eta}{2}g_{1}^{1}+u\alpha^{\ast}g_{1}^{2}\right) & -i\left( 
\frac{\eta}{2}g_{1}^{1}+u\alpha g_{1}^{2}\right) & -i\left( w_{1}-\mu\right)
& 0 & 0 \\ 
i\left( \frac{\eta}{2}g_{1}^{1}+u\alpha^{\ast}g_{1}^{2}\right) & i\left( 
\frac{\eta}{2}g_{1}^{1}+u\alpha g_{1}^{2}\right) & 0 & i\left(
w_{1}-\mu\right) & 0 \\ 
\vdots & \vdots & 0 & 0 & \ddots%
\end{array}
\right) ,   \label{eq:s-excitation matrix}
\end{equation}
\end{widetext}
and $\xi=\left( a_{in},a_{in}^{\dagger},0,0,\dots\right) ^{T}$. 
%give comment here, how do we produce the figure
% A phase transition occurs if $\mathcal{M}_{e}$ inherits a zero energy
% eigenstate. 
We numerically diagonalize $\mathcal{M}_{e}$ to obtain pairs of
% lowest 
atom polariton excitations $w_j^s$, with whose imaginary value approach $0$ we
determine the phase transition critical point, as shown in Fig.~\ref{fig:dissipation}(b).

In the normal phase $\alpha=0$, 
%$\left\{  \alpha= 0, \psi_{0} = (1, 0, 0, \dots)^{T}
%\right\} $,
for small $V_{l}$ one can expect that for the steady-state
 % $\psi$ have most
populations in the homogeneous mode and
 % a small excitation in 
 the mode of $%
\cos(2k_{l}x)$ are dominate, in the meanwhile the transformation matrix $\mathcal{O}$ is
essentially identity. Then one can roughly only account $j=f_{n+1}$ and $%
2f_{n}-f_{n+1}$ in $g_{j}^{1}$. %which have a relatively large
%and small value respectively.
Physically they correspond to the process in which the atoms in the
homogeneous mode and those in the mode of $\cos(2k_{l}x)$ are scattered by
the cavity photon, respectively. In this case we can consider the following
two block matrix 
\begin{equation}
\mathcal{M}_{ej}=\left( 
\begin{array}{cccc}
-i\Delta-\kappa & 0 & -i\frac{\eta}{2}g_{j}^{1} & -i\frac{\eta}{2}g_{j}^{1}
\\ 
0 & i\Delta-\kappa & i\frac{\eta}{2}g_{j}^{1} & i\frac{\eta}{2}g_{j}^{1} \\ 
-i\frac{\eta}{2}g_{j}^{1} & -i\frac{\eta}{2}g_{j}^{1} & -i\left(
w_{j}-\mu\right) & 0 \\ 
i\frac{\eta}{2}g_{j}^{1} & i\frac{\eta}{2}g_{j}^{1} & 0 & i\left(
w_{j}-\mu\right)%
\end{array}
\right) ,   \label{eq:s-reduced matrix}
\end{equation}
where $j=f_{n+1}$ or $2f_{n}-f_{n+1}$. $\mathcal{M}_{ej}$ can be
diagonalized and give 
\begin{equation}
\left( 
\begin{array}{cccc}
-i\Delta-\kappa & 0 & 0 & 0 \\ 
0 & i\Delta-\kappa & 0 & 0 \\ 
0 & 0 & -iw_{j}^{s} & 0 \\ 
0 & 0 & 0 & iw_{j}^{s}%
\end{array}
\right)   \label{eq:s-eigenvalue matrix}
\end{equation}
with $w_{j}^{s}\simeq\left( w_{j}-\mu\right) \sqrt{1-\left( \frac{\eta }{%
\eta_{c}}\right) ^{2}}$ %by considering that
under $|(w_{j}-\mu)/\Delta|<<1$, which indicate a critical pumping strength
of $\left( \eta_{c}g_{j}^{1}\right) ^{2}=\frac{\kappa^{2}+\Delta^{2}}{\Delta 
}\left( w_{j}-\mu\right) $. Beyond $\eta_{c}$ the quasiparticle
eigenfrequency $w_{j}^{s}$ would become complex and trigger instability of
the normal phase. For the Fibonacci sequence %case
of $\left( f_{n},f_{n+1}\right) =\left\{ \left( 2,3\right) ,\left(
3,5\right) \right\} $ considered in the main text the quasiparticle mode of $%
j=2f_{n}-f_{n+1}=1$ will have a smaller energy gap $w_{j}-\mu$ than that of $%
j = f_{n+1}$, resulting in a smaller $\eta_{c}$. This is the key mechanism
underlying a moiré lattice can stimulate the occurrence of phase
transition. %is determined from that.
In the absence of $V_{l}$, we have $g_{1}^{1}=\sqrt{2}$, $w_{1}-\mu=\omega
_{r}$ and $\Delta(\psi)=-\Delta_{c}$, resulting in a critical pumping $\sqrt{%
-(\Delta_{c}^{2}+\kappa^{2})\omega_{r}/2\Delta_{c}}$, which recover that of
the open Dicke model.

%give a discussion on the $M$ dependence here
With the increase of the moir\'e parameter $M$, the value of $w_{j} - \mu$
will decrease as can be seen from the expression of $\mathcal{C} (\alpha)$ (%
\ref{eq:s-scale parameter}) since $\omega_{k} = \omega_{r} / M^{2}$,
rendering that $\eta_{c}$ also decreases. Physically the behavior of critical pumping decreasing can be understood
as an effective enlargement of the lattice constant, which will decrease the
energy gap and thus facilitating phase transition.

%explain the role of v_l
%The incoherent cavity field excitation $\left\langle \delta a^{\dagger}\delta
%a\right\rangle $ is deduced in the following manner: First we derive the left
%eigenvector $v_{l}^{j}$ of the complex matrix $\mathcal{M}_{ej}$ with
%$v_{l}^{j}\mathcal{M}_{ej}=\lambda_{j}v_{l}^{j}$. Left multiply $v_{l}^{j}$ to
%Eq. (\ref{eq:s-langevin}) lead to%
%\[
%\frac{d}{dt}\mathcal{V}_{l}^{j}=
%\]

%Define $\mathcal{V}_{l}^{j} = v_{l}^{j} \mathcal{V}$

\section{Cavity spectrum}

\label{app_spectra}

The cavity field spectrum is defined as $S\left( \omega\right) =\mathcal{F}%
\left( \left\langle a^{\dagger}\left( t_{s}+t\right) a\left( t_{s}\right)
\right\rangle \right) $, where $\mathcal{F}\left( \cdot\right) $ indicate
Fourier transform with respect to $t$, $t_{s}$ is a time long enough for the
system reach stationary state. Since $\left\langle a^{\dagger}\left(
t_{s}+t\right) a\left( t_{s}\right) \right\rangle =N\left\vert
\alpha\right\vert ^{2}+\left\langle \delta a^{\dagger}\left( t_{s}+t\right)
\delta a\left( t_{s}\right) \right\rangle $, then we have%
\begin{align}
S\left( \omega\right) & =2\pi N\left\vert \alpha\right\vert ^{2}\delta\left(
\omega\right) +\int dte^{-i\omega t}\left\langle \delta a^{\dagger}\left(
t_{s}+t\right) \delta a\left( t_{s}\right) \right\rangle  \notag \\
& =2\pi N\left\vert \alpha\right\vert ^{2}\delta\left( \omega\right)
+\left\langle \delta a^{\dagger}\left( \omega\right) \delta a\left(
\omega^{\prime}\right) \right\rangle ,   \label{eq:s-spectrum1}
\end{align}
where $\delta a\left( \omega\right) $ represent the Fourier transform of $%
\delta a\left( t\right) $.

We left multiply $\mathcal{O}^{l}$ on both sides of Eq. (\ref{eq:s-langevin}%
), where the rows of $\mathcal{O}^{l}$ contain the left eigenvectors of the
complex matrix $\mathcal{M}_{e}$, from which one can have%
\begin{equation}
\frac{d}{dt}\mathcal{V}_{l}=\mathcal{WV}_{l}+\xi_{l}, 
\label{eq:s-left langevin}
\end{equation}
where $\mathcal{V}_{l}=\mathcal{O}^{l}\mathcal{V}=\left\{ \delta d,\delta
d^{\dagger},\delta e_{1},\delta e_{1}^{\dagger},\dots,\delta
e_{n_{c}},\delta e_{n_{c}}^{\dagger}\right\} ^{T}$, $\mathcal{W}$ is a
diagonal matrix with the diagonal elements $\left\{
-i\Delta-\kappa,i\Delta-\kappa,-iw_{1}^{s},iw_{1}^{s},%
\dots,-iw_{n_{c}}^{s},iw_{n_{c}}^{s}\right\} ^{T}$, and $\xi_{l}^{j}=%
\mathcal{O}_{j,1}^{l}a_{in}+\mathcal{O}_{j,2}^{l}a_{in}^{\dagger }$.
Introduce Fourier transforms to the quantum Langevin equations (\ref%
{eq:s-left langevin}) with%
\begin{align}
\mathcal{O}\left( \omega\right) =\frac{1}{\sqrt{2\pi}}\int_{-\infty}^{%
\infty}e^{i\omega t}\mathcal{O}\left( t\right) dt,
\nonumber\\
\mathcal{O}^{\dagger
}\left( -\omega\right) =\frac{1}{\sqrt{2\pi}}\int_{-\infty}^{\infty
}e^{i\omega t}\mathcal{O}^{\dagger}\left( t\right) dt,   \label{eq:s-fourier}
\end{align}
we can deduce $\mathcal{\tilde{V}}_{l}\left( \omega\right) =\left\{ \delta
d\left( \omega\right) ,\delta d^{\dagger}\left( -\omega\right) ,\delta
e_{1}\left( \omega\right) ,\delta e_{1}^{\dagger}\left( -\omega\right)
,\dots,\delta e_{n_{c}}\left( \omega\right) ,\delta e_{n_{c}}^{\dagger
}\left( -\omega\right) \right\} ^{T}$, with%
\begin{align}
\delta d\left( \omega\right) & =\frac{\tilde{\xi}_{l}^{1}}{-i\left(
\omega-\Delta\right) +\kappa},\delta d^{\dagger}\left( -\omega\right) =\frac{%
\tilde{\xi}_{l}^{2}}{-i\left( \omega+\Delta\right) +\kappa },  \notag \\
\delta e_{j}\left( \omega\right) & =\frac{\tilde{\xi}_{l}^{j+2}}{-i\left(
\omega-w_{j}^{s}\right) +\gamma},\delta e_{j}^{\dagger}\left( -\omega
\right) =\frac{\tilde{\xi}_{l}^{j+3}}{-i\left( \omega+w_{j}^{s}\right)
+\gamma},   \label{eq:s-fourier1}
\end{align}
where $\tilde{\xi}_{l}^{j}=\mathcal{O}_{j,1}^{l}a_{in}\left( \omega\right) +%
\mathcal{O}_{j,2}^{l}a_{in}^{\dagger}\left( -\omega\right) $, we have
phenomenologically introduced a decay rate $\gamma$ for the atomic polariton
modes.  % to account for their collision interactions with the surrounding
% Bogoliubov excitations.

We then perform a reverse transform $\mathcal{O}^{r}\mathcal{\tilde{V}}%
_{l}\left( \omega\right) $ with the columns of $\mathcal{O}^{r}$ contain the
right eigenvectors of $\mathcal{M}_{e}$, from which we can have%
\begin{align}
\delta a\left( \omega\right) &=\mathcal{O}_{11}^{r}\delta d\left(
\omega\right) +\mathcal{O}_{12}^{r}\delta d^{\dagger}\left( -\omega\right)
+\sum_{j=1}^{n_{c}}\left[ \mathcal{O}_{1,j+2}^{r}\delta e_{j}\left(
\omega\right) \right.
\nonumber\\
&\left.+\mathcal{O}_{1,j+3}^{r}\delta e_{j}^{\dagger}\left(
-\omega\right) \right] .   \label{eq:s-reverse}
\end{align}
Combining eq. (\ref{eq:s-fourier1}) and (\ref{eq:s-reverse}), we can have 
\begin{widetext}
\begin{align}
\left\langle \delta a^{\dagger}\left( \omega\right) \delta a\left(
\omega^{\prime}\right) \right\rangle & =\left\vert \mathcal{O}%
_{11}^{r}\right\vert ^{2}\left\langle \delta d^{\dagger}\left( \omega\right)
\delta d\left( \omega^{\prime}\right) \right\rangle 
% \nonumber\\
+\left\vert \mathcal{O}%
_{12}^{r}\right\vert ^{2}\left\langle \delta d\left( -\omega\right) \delta
d^{\dagger}\left( -\omega^{\prime}\right) \right\rangle  \notag \\
& +\sum_{j=1}^{n_{c}}\left[ \left\vert \mathcal{O}_{1,j+2}^{r}\right\vert
^{2}\left\langle \delta e_{j}^{\dagger}\left( \omega\right) \delta
e_{j}\left( \omega^{\prime}\right) \right\rangle 
% \right.
% \nonumber\\
% \left. 
+\left\vert \mathcal{O}%
_{1,j+3}^{r}\right\vert ^{2}\left\langle \delta e_{j}\left( -\omega\right)
\delta e_{j}^{\dagger}\left( -\omega^{\prime}\right) \right\rangle \right] 
\notag \\
& \simeq\sum_{j=1}^{n_{c}}\left[ \frac{\kappa\delta\left( \omega
-\omega^{\prime}\right) }{\left\vert -i\left( \omega-w_{j}^{s}\right)
+\gamma\right\vert ^{2}}\left\vert \mathcal{O}_{j+2,2}^{l}
% \right.
% \nonumber\\
% \left.
% &\times
\mathcal{O}%
_{1,j+2}^{r}\right\vert ^{2}+\frac{\kappa\delta\left( \omega-\omega^{\prime
}\right) }{\left\vert -i\left( \omega+w_{j}^{s}\right) +\gamma\right\vert
^{2}}\left\vert \mathcal{O}_{j+3,2}^{l}\mathcal{O}_{1,j+3}^{r}\right\vert
^{2}\right] ,   \label{eq:s-spectrum}
\end{align}
\end{widetext}
where we have assumed that the photonic polariton mode population $%
\left\langle \delta d^{\dagger}\left( \omega\right) \delta d\left(
\omega^{\prime}\right) \right\rangle $ is vanishingly small in the case of
large $\Delta_{c}$ ($\Delta$).

\section{Truncated Wigner methods}

\label{app_twa}

We adopt the truncated Wigner approximation (TWA) to study the dynamics and
obtain the results presented in Fig. 4 of the main text. In an open system,
we first apply Wigner-Weyl transform on both sides of 
%TWA states that the evolution equation for the Wigner function $\mathcal{W}$
%for a quantum state approximately follows the equation
%corresponding to
the master equation (1) %can be derived
to obtain the evolution equation for the Wigner function $\mathcal{W}$, in
which we neglect third-order derivatives under TWA. The resulting equation
is of the form of a Fokker-Planck equation, which can be simulated with
stochastic differential equations%
\begin{align}
\frac{d\alpha}{dt} & =\left[ i\left( \Delta_{c}-u\operatorname{Re}\left[ \psi^{\ast}%
\mathcal{M}_{2f_{n+1}}\psi\right] \right) -\kappa\right] \alpha
\nonumber\\
&-i\eta\operatorname{Re%
}\left[ \psi^{\ast}\mathcal{M}_{f_{n+1}}\psi\right] +\xi/\sqrt{N},  \notag \\
\frac{d\psi}{dt} & =-i\left[ \omega_{k}\mathcal{K}+u\left\vert
\alpha\right\vert ^{2}\left( 2\mathds{1}+\mathcal{M}_{2f_{n+1}}\right) +%
\frac{V_{l}}{4}\mathcal{M}_{2f_{n}}\right.
\nonumber\\
&\left.+\eta\operatorname{Re}(\alpha )\mathcal{M}%
_{f_{n+1}}\right] \psi,   
\label{eq:stochastic}
\end{align}
in which $\left\{ {\alpha(0),\ \psi(0)}\right\} $ are sampled from $\mathcal{%
W}(\alpha,\alpha^{\ast},\psi,\psi^{\ast},0)$ and $\xi$ is a complex Gaussian
random number with variance $\kappa$ and mean value $0$. Specifically, we
sample with%
\begin{equation}
\left( 
\begin{array}{c}
\alpha \\ 
\psi_{i}%
\end{array}
\right) =\left( 
\begin{array}{c}
\left( m+in\right) /2\sqrt{N} \\ 
o_{i}+\left( p_{i}+iq_{i}\right) /2\sqrt{N}%
\end{array}
\right) ,   \label{eq:sample}
\end{equation}
where $m,n,p_{i},q_{i}$ are independent random numbers drawn from a Gaussian
distribution with zero mean and unit variance. It means that initially
cavity is in coherent vacuum. For an initial state with all atoms in $%
\psi_{0}$ while the other atomic modes are left empty, we have $o_{0} = 1$
and $o_{i} = 0$ otherwise. In practice, we sample a system of $N=1000$ with $%
500$ trajectories. 
%whether we need to compare twa with monte-carlo and say they are in good agreement

\section{Atom diffusion dynamics}

\label{app_diff}

We have adopted two different schemes to study atomic diffusion and the
physics discovered by both schemes are in qualitative agreement. 
% with each other. 
In both schemes the cavity field is treated using TWA with the details
described in \ref{app_twa}. Difference lies in that in one scheme the atomic
wavefunction is simulated with Gross-Pitaevskii equation%
\begin{align}
i\frac{d\Psi\left( x,t\right) }{dt}&=\left[ -\frac{\hbar^{2}}{2m}%
\partial_{x}^{2}+4u\left\vert \alpha\right\vert ^{2}\cos^{2}\left(
k_{c}x\right) +V_{l}\cos^{2}\left( k_{l}x\right) \right.
\nonumber\\
&\left.+2\eta\operatorname{Re}%
(\alpha)\cos\left( k_{c}x\right) \right] \Psi\left( x,t\right) , 
\label{gp equation}
\end{align}
in which $\left\vert \alpha\right\vert ^{2}$ and $\operatorname{Re}(\alpha)$ are
taken from ensemble average of TWA simulation on cavity field. We use this
scheme to generate results presented in the main text.

% The atomic diffusion is studied . 
In the other scheme we apply mode expansion to the atomic wavefunction and
simulate its dynamics using TWA. For atoms initially prepared in a Gaussian
wavepacket with width $w$%
\begin{equation}
\phi\left( x\right) =\left( w\sqrt{2\pi}\right) ^{-1/2}\exp\left[ -\frac{%
\left( x-L/2\right) ^{2}}{4w^{2}}\right] ,   \label{eq:initial gaussian}
\end{equation}
it can be represented by the $1$-periodic Fourier series as%
\begin{align}
\phi\left( x\right) &=a_{0}\sqrt{\frac{1}{L}}+\underset{N\rightarrow\infty }{%
\lim}\sqrt{\frac{2}{L}}\left( \sum \limits_{n=1}^{N}a_{n}\cos\left( 2\pi
nx\right) \right.
\nonumber\\
&\left. +\sum \limits_{n=1}^{N}b_{n}\sin\left( 2\pi nx\right) \right) , 
\label{eq:fourier expand}
\end{align}
where%
\begin{align}
a_{0}&=\sqrt{\sqrt{2\pi}2w}\operatorname{erf}\left( \frac{1}{4w}\right) ,
\nonumber\\
a_{n}&=2\sqrt{%
\sqrt{2\pi}w}\cos\left( n\pi\right) e^{-\left( 2n\pi w\right) ^{2}}\operatorname{Re}%
\left[ \operatorname{erf}\left( \frac{1}{4w} \right.\right.
\nonumber\\
&\left.\left. +2in\pi w\right) \right] , 
\label{eq:fourier coefficients}
\end{align}
with $\operatorname{erf}(\cdot)$ the error function. Compare (\ref{eq:fourier expand}%
) with the mode expansion (3) in the main text, in TWA sampling (\ref%
{eq:sample}) we choose $o_{i} = a_{i}$.

%figure of gaussian wavepacket=====
\begin{figure}[htb]
\centering
\includegraphics[width=3 in]{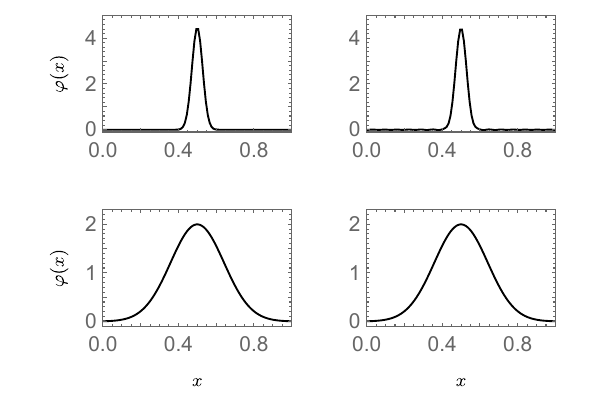} 
\caption{ Original atomic wavepacket (left column) compared with those from
reconstruction (right column). The top row is for $w = 0.02$, while the
second row is for $w = 0.1$. }
\label{fig:gauss}
\end{figure}

Note that as we have omitted the sine term and truncated at $n_{c}$, these
will affect the atomic wavepacket reconstructed after TWA simulation,
especially when the wavepacket width is small. This is indicated in Fig.~\ref%
{fig:gauss}, in which we compared the original wavepacket with the
reconstructed one. However this only cause quantitative differences and will
not affect our main results in the text. %\newpage
%%\bibliographystyle{apsrev4-1}

%comment on the diffusion
With periodic boundary condition considered here, 
% the atomic wavepacket donot
% spread continuously. Instead, 
as shown in Fig.~\ref{fig:diff}(b), the width increases at initial time and
then gradually approaches a steady value. In the meanwhile, the cavity
photon number jump from zero before reaching saturation. We plot the steady
state width $w_{s}$ versus the pumping strength in Fig.~\ref{fig:diff}(a). 
%for
%why width at $M = 3$ is smaller than $5$.
As expected, since the lattice constant 
%of the external lattice at $M = 5$ is larger than that at $M = 3$,
becomes larger with increased $M$, %rendering
one will have broader wavepacket in the $M = 5$ case as compared with $M = 3$
for an initial identical wavepacket. This is true at small $\eta$. 
%despite the initial identical wavepacket.
$w_{s}$ abruptly decreases upon the pumping strength passing across a
critical value. This diffusion behavior tuned between two distinct diffusive
regimes arises from superradiant phase transitions: (i) For small pumping
strengths, cavity field is not excited and atomic wavepacket spreads until
it reach the unit cell boundary, with the width showcasing %quasi-periodic
damped oscillation as illustrated by the green line in Fig.~\ref{fig:diff}%
(b). In the steady state, atomic wavefunction distribution is localized in
momentum space as shown in the upright inset of Fig.~\ref{fig:diff}(a),
indicating extended distribution in coordinate space. (ii) While at large
pumping, intracavity field build up to coform a moir\'{e} lattice. Different
atomic momentum modes are populated via scattering, resulting in momentum
space extended distribution at steady state (downleft inset of Fig.~\ref%
{fig:diff}(a)), and henceforce suppressed expansion and width.

%Fig.~\ref{fig:diff}(a) also exemplify that moir\'{e} parameter can be used to control the diffusion behavior.

%fig-diffusion present the quantum diffusion results===========
\begin{figure}[htb]
\includegraphics[width=3 in]{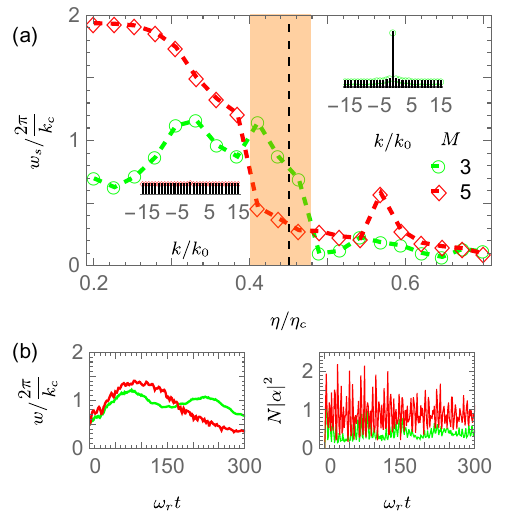} %
\caption{ Quantum diffusion dynamics calculated with TWA. (a) Atom
wavepacket width at steady state $w_{s}$ versus pumping strength $\protect%
\eta$. In the shaded region $w_{s}$ at $M = 3$ is larger than that at $M = 5$%
, with the insets demonstrating localized (extended) atomic momentum
distribution for $M = 3$ ($5$) respectively at a $\protect\eta$ marked by
the black dashed line. The corresponding evolution dynamics of atomic
wavepacket width $w$ and cavity photon number are given in (b). The
parameters are the same as those in the main text. }
\label{fig:diff}
% \centering
%be careful on the
%scale, this figure give the phase transition, mode occupation and excitation
%number
\end{figure}
%==================================

Noteworthy that in the shaded region of Fig.~\ref{fig:diff}(a), for the very
same pumping strength, one counter-intuitively have larger $w_{s}$ in the $M
= 3$ case instead of $M = 5$. In this specific parameter region, atomic
wavepacket diffuse in distinct different manner determined by the moir\'{e}
parameter: In the %a moir\'{e} lattice with
$M = 3$ case atomic diffusion %in a manner
behavior can be depicted by (i), while the atomic diffusion in the $M = 5$
case %can be
is attributed to (ii). This completely opposite behavior is %aroused
manipulated solely by a change in %the moir\'{e} parameter
$M$ with all the other physical parameters identical, thus can be regarded
as moir\'{e} effect. %This phenomenon originate from the
%dependence of superradiant phase transition on the
%moir\'{e} superradiant phase transition,
%and can be explained with
One can understand this phenomenon as the result of interplay between
quantum interference (Anderson localization) and quantum fluctuations of the
cavity field. %note that this behavior cannot have without the cavity
%without the cavity
The aid of cavity is indispensable to observe the moir\'{e} effect.

\end{document}